\documentclass[showpacs,preprintnumbers,amsmath,amssymb,prb,twocolumn,floatfix]{revtex4}
\usepackage[dvips]{graphics,color}
\usepackage{dcolumn}
\begin{document}
\title{Influence of a Transport Current on Magnetic Anisotropy in Gyrotropic Ferromagnets}
\author{Ion Garate, A.H. MacDonald}
\affiliation{Department of Physics, The University of Texas at Austin, Austin, Texas 78712}
\date{\today}
\begin{abstract}
Current-induced torques are commonly used to manipulate non-collinear 
magnetization configurations.
In this article we 
discuss current-induced torques present in a certain class of collinear magnetic systems,   
relating them to current-induced changes in magnetic anisotropy energy.  
We present a quantitative estimate of their characteristics in uniform strained ferromagnetic (Ga,Mn)As.
\end{abstract}
\maketitle

\section{Introduction}
The interplay between transport currents and magnetization dynamics continues to be a major research topic in 
ferromagnetic metal spintronics.\cite{reviews}  The current understanding of this 
class of phenomena has been derived mainly 
from numerous studies of spin-transfer torques (STTs), which arise when spin polarized currents traverse {\em non-collinear} magnetic systems. 
STTs can be exploited to achieve current-induced magnetization reversal and current-induced domain-wall motion, both of which have 
potentially important technological applications.

There have been comparatively few studies of the influence of transport currents on magnetization in {\em uniform} ferromagnets, presumably because spin transfer torques vanish in these systems. Yet, as pointed out independently by several researchers\cite{manchon, chernyshov, our abstract}, current-induced reorientation of magnetization does occur in 
some uniform ferromagnets.  The first experimental fingerprint of this phenomenon was uncovered by Chernyshov {\em et al.}\cite{chernyshov} who demonstrated that an electric current 
alters magnetization reversal characteristics in strained (Ga,Mn)As films with a single magnetic domain. 
    
STTs can be considered to be one member of a family
of current-induced torque (CIT) effects by which 
transport currents influence magnetization in ferromagnetic or antiferromagnetic\cite{CITAF}
systems.  The aim of this paper is to contribute to the theoretical analysis of current-induced torques in uniformly magnetized ferromagnets. 

In Sec. II we study the effect responsible for this type of torque, which we 
refer to as the ferromagnetic inverse spin-galvanic effect.\cite{edelstein, ganichev}
In non-magnetic conductors the inverse spin-galvanic effect (ISGE) refers to 
current-induced spin density.  Since a non-zero spin-density already appears in the 
equilibrium state of a ferromagnet, the ferromagnetic inverse spin-galvanic effect 
has a distinct experimental signature. 
Specifically, we find that in gyrotropic ferromagnets the magnetization direction is 
altered by a steady-state transport current.
At a conceptual level, we associate this reorientation with a change in magnetic anisotropy in the presence of a transport current. An important implication of this connection is that the magnetic anisotropy energy in the transport steady state of a 
ferromagnet which exhibits the ISGE is {\em not} invariant under magnetization reversal, essentially because the applied current breaks time reversal invariance.
At a practical level, we provide a concise analytical expression for the current-induced change in the magnetic anisotropy. This expression is suitable for evaluation from first principles because it requires the knowledge of only the band structure of the ferromagnet and the lifetime of the Bloch states. At a technical level, our theory allows for the spatial inhomogeneities that inevitably occur in the {\em magnitude} of the ferromagnet's exchange field at atomic lenghtscales. 

In Sec. III we carry out quantitative calculations for the ISGE
of strained (Ga,Mn)As using a 4-band Kohn-Luttinger model. 
This calculation directly 
addresses the experiment by Chernyshov {\em et al.}\cite{chernyshov} and corroborates their interpretation of the data. By computing the anisotropy field both in absence and in presence of an electric current, we find that in (Ga,Mn)As magnetization 
reversal may in principle be achieved solely by electric means: the required critical current densities are in the order of $10^6-10^7 {\rm A/cm^2}$ and depend on the strain, Mn concentration and hole density. Sec. IV contains a brief summary and presents our conclusions.

The main conclusions of our work coincide with those reached by Manchon and Zhang in their independent and 
previously published work described in Ref.~[\onlinecite{manchon}]. 
Yet, our analysis highlights aspects that have not been emphasized previously. 
First, we assert that in ferromagnets with inversion symmetry,
the current-induced spin-density vanishes {\em to all orders} in the strength of 
the spin-orbit interaction. 
Second, when evaluating the current-induced spin polarization we include a contribution from interband coherence
which can become quantitatively important in disordered ferromagnets such as (Ga,Mn)As. 
Third, we identify the current-induced transverse spin-density associated with 
the ISGE in ferromagnets as a consequence of 
a change in magnetic anisotropy in the 
presence of an electric current. 
We thus promote transport currents to the same status as temperature\cite{ranvah}, gate voltages\cite{chiba, maruyama, weiler}, strain\cite{botters, lemaitre}  and chemical processes\cite{gambardella}, all of which are well-established control parameters for the tuning of magnetic anisotropy.  
 
\section{Theory of the Ferromagnetic Inverse Spin-Galvanic Effect}

In non-magnetic metals or semiconductors that are gyrotropic, {\em i.e.} non-centrosymmetric {\em and} chiral,\cite{inversion} a DC charge current is 
generically accompanied by a non-zero spin polarization.\cite{edelstein} This phenomenon is sometimes referred to as the inverse spin galvanic effect (ISGE).\cite{ganichev}  
Because of the advent of spintronics and subsequent attempts to control spin polarization by electric means, 
even in paramagnetic materials, the ISGE has received 
widespread experimental\cite{cisp experiment} and theoretical\cite{cisp theory} attention.
The ISGE is purely a consequence of symmetry since i) current, which is odd under   
time reversal, is the dissipative response of a conductor to a DC electric-field,
ii) spin is also odd under time reversal and therefore allowed as part of the dissipative response, and  
iii) axial vectors (like spin) and polar vectors (like current)
are coupled in gyrotropic materials.\cite{gyrotropy def}   
The direction of the carriers' spin is determined by the direction of the electric field as well as by the axis along which inversion symmetry is broken.\cite{symmetry arguments} 

The ISGE is sometimes viewed as a possible route toward the development of spintronics effects in 
paramagnetic materials that are as robust  
as effects like giant magnetoresistance that occur 
only in ferromagnetic materials.  Partly because spin-orbit interactions tend to be 
fairly weak, it appears to be difficult to make spin-galvanic effects in normal metals useful.   
In this section we turn the tables on this strategy by concentrating on the inverse spin-galvanic effect in {\em magnetic} conductors. 

In uniformly magnetized ferromagnets with inversion symmetry, the transport current is spin polarized 
because the conductivities of majority and minority spin channels are different.
This familiar fact is unrelated to the ISGE. 
Since spin-polarization is already present in the thermodynamic equilibrium state of a 
ferromagnet, the ferromagnetic ISGE is manifested not by the presence of a non-zero spin-density 
but instead by a change in magnetization direction in the non-equilibrium steady-state which is dependent on the magnitude 
and direction of the electric field.  
In this paper we formulate a theory of the ISGE in ferromagnets by evaluating the torque which acts on the 
collective magnetization of a magnetic conductor due to spin-orbit interactions in the presence of a transport current.
When the current is set to zero, the torque we evaluate vanishes along easy (and hard) magnetization
directions and is normally viewed\cite{Stohr_Magnetism} as a precessional torque due to magnetocrystalline anisotropy fields.
These torques are in turn associated with the magnetization-direction dependence of the magnetocrystalline anisotropy energy.
At zero current, the anisotropy torques must change sign when the magnetization direction is reversed because time 
reversal symmetry requires that the anisotropy energy be invariant under reversal.
The ferromagnetic ISGE in gyrotropic crystals may be viewed as a change in anisotropy torque due to a transport current.
Significantly, the ISGE torques are {\em not} odd under magnetization reversal.


The ferromagnetic ISGE is reminiscent of the magnetoelectric phenomena that 
have been extensively studied in multiferroic materials,\cite{multiferroics} 
{\em i.e.} materials in which magnetism coexists with ferroelectricity. 
A common characteristic of multiferroic perovskites is the presence of canted magnetism that stems from the Dzyaloshinskii-Moriya interaction. 
Since the direction of canting is determined by the symmetry of the crystal, one can envisage\cite{ederer} scenarios 
in which an electric-field-mediated reversal of the ferroelectric polarization causes a simultaneous reversal of the canting angle or of the magnetization. 
Another interesting property of multiferroic materials is the coupling between ferroelectricity and antiferromagnetism.\cite{zhao} This 
coupling makes it possible to switch the magnetization of an exchange-biased ferromagnet by the application of an electric field.
In spite of the contextual similarities, there are fundamental differences between the aforementioned phenomena and the ferromagnetic ISGE. 
For one thing, ferroelectricity occurs only in insulators while the ISGE occurs only in conductors.

\begin{figure}
\begin{center}
\scalebox{0.4}{\includegraphics{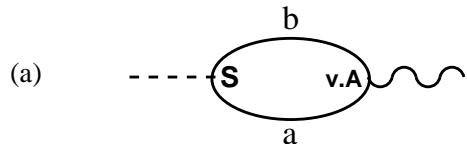}}
\caption{Feynman diagram that encodes the transverse spin density induced by a current (ferromagnetic ISGE effect)
which results in a change in the steady-state magnetization direction. $a$ and $b$ are band labels for the quasiparticle and the quasihole.}
\label{fig:bubbles}
\end{center}
\end{figure}

We evaluate the ferromagnetic ISGE microscopically within the framework of linear response theory (Fig. (~\ref{fig:bubbles}a)):
\begin{equation}
\delta s^i=\chi^{i,j}_{S,E} E^{j},
\end{equation}
where $\delta s^i$ is the current-induced spin density ($i\in\{x,y,z\}$), ${\bf E}$ is the applied electric field, and
$\chi$ the dissipative spin-current response function: 
\begin{equation}
\label{eq:chi}
\chi^{i,j}_{S,E}=\frac{1}{2\pi}{\rm Re}\sum_{{\bf k},a,b} s^i_{a,b}({\bf k}) v^j_{b,a}({\bf k}) \left(G_{{\bf k},a}^R G_{{\bf k},b}^A-G_{{\bf k},a}^R G_{{\bf k},b}^R\right).
\end{equation}
This linear response theory expression applies for time-independent uniform applied electric fields, and may be derived in the standard way\cite{mahan} by analytically continuing the imaginary\cite{diamagnetic} part of $\lim_{\omega\to 0}\tilde{\chi}_{S,E}^{i,j}/\omega$, where 
\begin{equation}
\label{eq:chi0}
\tilde{\chi}_{S,E}^{i,j}=-T\sum_{i\omega_n{\bf k},a,b} s^i_{a,b}({\bf k}) v^j_{b,a}({\bf k}) G_{{\bf k},a}(i\omega_n) G_{{\bf k},b}(i\omega_n+i\omega),
\end{equation}
$\omega_n=(2n+1)\pi T$ is the Matsubara frequency at temperature $T$ and $\omega$ is the frequency of the external field. 
In Eq.(~\ref{eq:chi}) $s^i_{a,b}({\bf k})$ and $v^j_{b,a}({\bf k})$ are the ${\bf k}$-dependent matrix-elements of the spin and velocity operators
($O_{a,b}({\bf k})\equiv \langle a,{\bf k}|O|b,{\bf k}\rangle$) between Bloch states ($|a,{\bf k}\rangle$) in bands $a$ and $b$.
Note that the Bloch states are in general spinors in which orbital and spin degrees of freedom are entangled. $G_{{\bf k},a}^{R(A)}=1/(\epsilon_F-\epsilon_{{\bf k},a}+(-) i/2\tau_{{\bf k},a})$ is the retarded (advanced) Green's function evaluated at the Fermi energy $\epsilon_F$, and $\tau_{{\bf k},a}$ is the quasiparticle lifetime.
For simplicity we have ignored disorder vertex corrections to both velocity and spin operators. 
In the numerical calculations discussed in Section III
we will in addition take the quasiparticle lifetime to be a phenomenological parameter which is 
independent of momentum and band labels.  

As we discuss below, the transverse components of the spin-density are directly related to the anisotropy field, which exerts a torque on the macrospin. On the same footing, the current-induced contribution to the transverse spin density is directly related to the current-induced contribution to the anisotropy field.


For a ferromagnet with inversion symmetry $\chi_{S,E}=0$ irrespective of spin-orbit interaction strength, for essentially
the same reasons as the ISGE vanishes in normal conductors with inversion symmetry.\cite{liu} 
This property can be verified by recognizing that in presence of inversion symmetry the Hamiltonian of the ferromagnet is invariant under ${\bf k}\to -{\bf k}$, which implies that $G_{\bf k}=G_{- \bf k}$, $s_{a,b}({\bf k})=s_{a,b}(-{\bf k})$ and $v_{a,b}({\bf k})=-v_{a,b}(-{\bf k})$. Consequently, the right hand side of Eq. ~(\ref{eq:chi}) vanishes after summing over all ${\bf k}$. 
From a crystal symmetry classification standpoint there are 21 non-centrosymmetric crystal classes, among which three ($T_d$, $C_{3h}$ and $D_{3h}$) are not gyrotropic.
The occurence of the ISGE is therefore restricted to 18 crystal classes.\cite{ganichev} 
 
The main objective of this section is to relate the ferromagnetic ISGE to a current-induced change in the magnetic anisotropy field, yet before we do so it is beneficial to pave the way by reviewing the nuances of magnetic anisotropy in electric equilibrium. 
In the absence of currents, magnetic anisotropy describes the dependence of the free energy of a ferromagnet on the direction of its magnetization.\cite{vonsovski} 
Magnetic anisotropy originates from\cite{ma reviews} magnetic dipolar interactions and spin-orbit interactions. 
The former lead to shape anisotropy in non-spherical samples 
while the latter produce magnetocrystalline anisotropy by communicating the lack of rotational symmetry in the crystalline lattice to
the spin degrees of freedom. 
In practice, magnetic anisotropy reveals itself in dynamical processes such as ferromagnetic resonance through an anisotropy field that forces the magnetization to precess 
unless it is along an easy or hard axis, {\em i.e.} along a direction in which the anisotropy energy is minimized or maximized. 
This precessional magnetization dynamics is properly characterized by the Landau-Lifshitz equation, $\partial_t\hat\Omega=\hat\Omega\times H_{\rm eff}$, where $\hat\Omega$ is the direction of the 
ferromagnet's collective dynamical variable (which may be chosen to be either the magnetization or the ferromagnetic exchange field) 
and $H_{\rm eff}$ is an effective magnetic field, taken here to 
include reactive as well as dissipative processes.\cite{fanhle,alpha1} 
The anisotropy field may then be defined as the contribution to the non-dissipative part of the effective magnetic field which survives in the absence of true magnetic fields:
\begin{equation}
{\bf H}_{\rm an}=-\frac{1}{S_0}\frac{\partial E_{GS}}{\partial\hat\Omega},
\end{equation}
where $E_{\rm GS}$ is the ground state energy of the ferromagnet in equilibrium (we take zero temperature throughout) and $S_0$ is the total spin (magnetization$\times$ volume) of the ferromagnet. 

\begin{figure}
\begin{center}
\label{fig:cartoon}
\scalebox{0.4}{\includegraphics{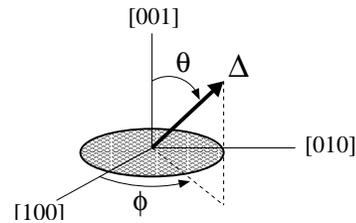}}
\caption{Cartoon of a magnetic thin film (shaded area). The exchange field ${\bf \Delta}$ is an effective 
magnetic field which is parallel to the magnetization only when it points along easy or hard crystalline directions.
The orientation of ${\bf \Delta}$ can be specified by the polar and azimuthal angles $\theta$ and $\phi$.
The relationship between the direction of ${\bf \Delta}$ and the direction of magnetization is altered
by an electric current in gyrotropic ferromagnets.} 
\end{center}
\end{figure}

When we discuss (Ga,Mn)As in the following section, we will use 
spherical coordinates (Fig. ~\ref{fig:cartoon}) in which the anisotropy field may be written as
\begin{equation}
\label{eq:spherical}
{\bf H}_{\rm an}=H_{\phi}\hat\phi+H_{\theta}\hat\theta,
\end{equation}
where $\hat\phi$ and $\hat\theta$ are the azimuthal and the polar unit vectors, respectively. The longitudinal component of the anisotropy
field is irrelevant because $\hat\Omega\times\hat\Omega=0 $.

In order to elaborate on the microscopic theory of the anisotropy field in a concrete way 
we work within the spin-density-functional theory of a magnetic material, 
in which the effective Hamiltonian that describes the theory's Kohn-Sham 
quasiparticles can be expressed as 
\begin{equation}
\label{eq:H_KS}
{\cal H}={\cal H}_{\rm kin}+{\cal H}_{\rm so}-{\bf \Delta}\cdot{\bf s}.
\end{equation}
In Eq. ~(\ref{eq:H_KS}) ${\bf\Delta}=\Delta_0({\bf r}) \hat\Omega$ is the exchange effective-magnetic-field of the ferromagnet, $\hat\Omega$ is the direction of the exchange field,
${\bf s}$ is the quasiparticle spin operator, ${\cal H}_{\rm so}$ captures spin-orbit interactions, and ${\cal H}_{\rm kin}$ collects all spin-independent terms in the Kohn-Sham Hamiltonian. 
In this work we characterize the macrostate of a ferromagnet by specifying the direction of the exchange field. $\hat\Omega$ is assumed to be uniform in space but the magnitude $\Delta_0({\bf r})$ of the exchange field is allowed to have spatial dependence at the atomic lengthscale.\cite{alpha1} 
We neglect dipolar interactions since they are not directly influenced by currents 
and can normally be cleanly separated from magnetocrystalline anisotropy. 

It follows that the zero-temperature anisotropy field is given by
\begin{equation}
\label{eq:H_an_eq}
{\bf H}_{\rm an}=-\frac{1}{S_0}\sum_{{\bf k},a} \frac{\partial\epsilon_{{\bf k},a}}{\partial\hat\Omega} f_{{\bf k},a}.
\end{equation}
In Eq. ~(\ref{eq:H_an_eq}) we have used\cite{force theorem} $E_{\rm GS}=\sum_{{\bf k},a}\epsilon_{{\bf k},a} f_{{\bf k},a}$, where $\epsilon_{{\bf k},a}$ is the energy of the Bloch state quasiparticles and $f_{{\bf k},a}=\Theta(\epsilon_F-\epsilon_{{\bf k},a})$ is the equilibrium occupation factor at zero temperature. Furthermore we have exploited the fact that 
\begin{equation}\sum \epsilon_{{\bf k},a} \frac{\partial f_{{\bf k},a}}{\partial\hat\Omega}\simeq \epsilon_F \sum_{{\bf k},a} \frac{\partial f_{{\bf k},a}}{\partial\hat\Omega}=0,
\end{equation}
since the number of electrons in the ferromagnet is invariant under rotations of the magnetization. 
This implies a $\hat\Omega$-dependence of the Fermi energy,\cite{daalderop, wunderlich} which is taken into account in the calculations of Sec. III. 

Eq.~(\ref{eq:H_an_eq}) may be rewritten in a more informative manner using the Feynman-Hellmann theorem, which implies that  
\begin{equation}
\label{eq:FH}
\frac{\partial \epsilon_{{\bf k},a}}{\partial\Omega_i}=\langle a,{\bf k}|\frac{\partial {\cal H}}{\partial\Omega_i}|a, {\bf k}\rangle = -\langle a,{\bf k}|\Delta_0({\bf r}) s_i|a, {\bf k}\rangle. 
\end{equation}
Then,
\begin{equation}
\label{eq:H_an_eq2}
{\bf H}_{\rm an}=\frac{1}{S_0}\sum_{{\bf k},a} \langle a,{\bf k}|\Delta_0({\bf r}){\bf s}| a,{\bf k}\rangle f_{{\bf k},a},
\end{equation}
where $\langle a,{\bf k}|\Delta_0({\bf r}){\bf s}| a,{\bf k}\rangle\equiv\int d{\bf r} \Delta_0({\bf r})\langle a,{\bf k}|{\bf r}\rangle {\bf s} \langle {\bf r}|a,{\bf k}\rangle$. 

For the envelope-function model we use in the next section, 
the magnitude of the exchange field is a spatially constant $\Delta_0$ and the torque exerted by the anisotropy field is simply equal to the $\Delta_0$ times the 
transverse spin-density divided by the total spin of the ferromagnet.
In {\em ab initio} calculations, the magnitude of the exchange field always varies 
substantially on an atomic scale and, as we have emphasized previously,\cite{alpha1} this variation must 
be accounted for.  In this case the anisotropy field is evaluated by integrating the 
product of the exchange field magnitude and transverse spin density over space.

Eq.~(\ref{eq:H_an_eq2}) may be separated into azimuthal and polar components:
\begin{eqnarray}
\label{eq:H_pt}
H_{\phi}&=& \frac{1}{S_0}\sum_{{\bf k},a} \langle a,{\bf k}|\hat{z}\cdot({\bf\Delta}\times{\bf s})|a,{\bf k}\rangle\nonumber\\
H_{\theta}&=& \frac{1}{S_0}\sum_{{\bf k},a} \langle a,{\bf k}|\hat{\phi}\cdot({\bf\Delta}\times{\bf s})|a,{\bf k}\rangle
\end{eqnarray}
If we neglect spatial variations of $\Delta_0({\bf r})$, Eqs.~(\ref{eq:H_an_eq2}) and ~(\ref{eq:H_pt}) indicate that the torque created by the anisotropy field will vanish when the (spin) magnetization $\sum\langle{\bf s}\rangle f$ is parallel to the exchange field. Conversely, whenever the direction of magnetization is misaligned with ${\bf \Delta}$, the anisotropy field will be nonzero and will produce a torque on the magnetization. In transition metals spin-orbit interactions produce a misalignment between the exchange field and the magnetization, unless $\hat\Omega$ is pointing along some special crystalline direction that corresponds (by definition) to an easy or hard axis. A similar picture applies to local-moment ferromagnets as well, where due to spin-orbit coupling the direction of the local moments is generally misaligned with the direction of the itinerant spin density.

One of the targets of this section is to present formulae that are useful for researchers working on both model systems as well as {\em ab-initio} electronic structure calculations. Therefore, we digress to explain that Eq.~(\ref{eq:H_an_eq2}) is equivalent to the alternative expressions found in {\em ab-initio} studies.  In first principles magnetic anisotropy theory\cite{daalderop,wang} Eq. ~(\ref{eq:FH}) 
has been approached from a different perspective. 
In such approach it is customary to choose the spin quantization axis along the direction of magnetization, so that ${\bf \Delta}\cdot{\bf s}\equiv\Delta_0 s_z$ is independent of $\hat\Omega$. 
When this choice is made, the spin-orbit term in the Hamiltonian becomes explicitly $\hat\Omega$-dependent. Consequently,
\begin{equation}
\label{eq:so}
\frac{\partial{\epsilon_{{\bf k},a}}}{\partial\hat\Omega}=\langle a,{\bf k}|\frac{\partial {\cal H}_{\rm so}}{\partial \hat\Omega}|a, {\bf k}\rangle.
\end{equation}
The anisotropy field is then evaluated combining Eq.~(\ref{eq:so}) with 
the force theorem\cite{force theorem} and a full-potential electronic-structure calculation.\cite{wang} 
Of course, the final result is invariant with respect to the choice of the spin quantization axis. In order to prove the equivalence of Eqs.~(\ref{eq:FH}) and ~(\ref{eq:so}) it is convenient to rewrite\cite{stiles}  Eq.~(\ref{eq:so}) as $\partial\epsilon/\partial\phi=\langle\partial_{\phi}[ \exp(i {\bf s}\cdot\hat{z}\phi) {\cal H}_{\rm so} \exp(-i {\bf s}\cdot{\hat z} \phi)]\rangle\vert_{0}$ and $\partial\epsilon/\partial\theta=\langle\partial_{\theta}[\exp(i {\bf s}\cdot\hat\phi\theta) {\cal H}_{\rm so} \exp(-i {\bf s}\cdot\hat\phi\theta)]\rangle\vert_{0}$. 
To see that these expressions agree with Eq.~(\ref{eq:H_pt}) note that $[{\cal H}_{\rm so},{\bf s}]=[{\cal H}-{\cal H}_{\rm kin}+{\bf\Delta}\cdot{\bf s},{\bf s}]$, 
that $[{\cal H}_{\rm kin},{\bf s}]\equiv 0$, and that $\langle a,{\bf k}|[{\cal H},{\bf s}]|a,{\bf k}\rangle=(\epsilon_{{\bf k},a}-\epsilon_{{\bf k},a})\langle a,{\bf k}|{\bf s}|a,{\bf k}\rangle =0$. 
In this way the derivative of energy with respect to magnetization direction can be related to the exchange term in the 
Kohn-Sham equation rather than to the spin-orbit coupling term.  Eqs.~ (\ref{eq:H_an_eq2}) and ~(\ref{eq:H_pt}) are recovered after using 
$[s_i,s_j]=i\epsilon_{ijk} s_k$ to simplify $\langle [{\bf \Delta} \cdot{\bf s},{\bf s}]\rangle$. 


We now show that the Green's function expression we use to evaluate the ferromagnetic ISGE (Eq.(~\ref{eq:chi0}))
corresponds to the current-induced change in Eq.~ (\ref{eq:H_an_eq2}).
We begin by mentioning that the application of an electric current can alter the magnetic anisotropy field, which leads to a current-induced torque on the magnetization.
For an arbitrary orientation of the exchange field, the change is given by
\begin{widetext} 
\begin{equation}
\label{eq:d_H_an}
{\bf \delta H}_{\rm an}=\frac{1}{S_0} \sum_{{\bf k},a}\delta(\langle a,{\bf k}|\Delta_0({\bf r}){\bf s}| a,{\bf k}\rangle) f_{{\bf k},a} + \frac{1}{S_0}\sum_{{\bf k},a}\langle a,{\bf k}|\Delta_0({\bf r}){\bf s}|a, {\bf k}\rangle\delta f_{{\bf k},a}.
\end{equation}
\end{widetext} 
Adopting the relaxation-time approximation,  $\delta f$ reads 
\begin{equation}
\label{eq:occupation}
\delta f_{{\bf k},a}={\bf E}\cdot {\bf v}_{a,a} \, \frac{\partial f_{{\bf k},a}}{\partial\epsilon_{{\bf k},a}} \, \tau_{{\bf k},a},
\end{equation}
and for the change in the matrix elements we use 
\begin{equation}
\label{eq:state}
\delta (\langle a,{\bf k}|\Delta_0{\bf s}|a,{\bf k}\rangle)=\langle a,{\bf k}|\Delta_0 {\bf s}\delta(|a,{\bf k}\rangle) + {\rm c.c}
\end{equation}
with
\begin{equation}
\label{eq:state2}
\delta(|a,{\bf k}\rangle)=\frac{e^{i\omega t}}{i\omega}\sum_{b\neq a}|b,{\bf k}\rangle\frac{\langle b,{\bf k}|{\bf v}\cdot{\bf E}|a,{\bf k}\rangle}{\epsilon_{{\bf k},a}-\epsilon_{{\bf k},b}+\omega} +(\omega\to -\omega).
\end{equation}
In Eq.~(\ref{eq:state2}) we have once again appealed to linear response theory and have used the fact that the electric field is uniform.

\begin{figure}
\begin{center}
\scalebox{0.4}{\includegraphics{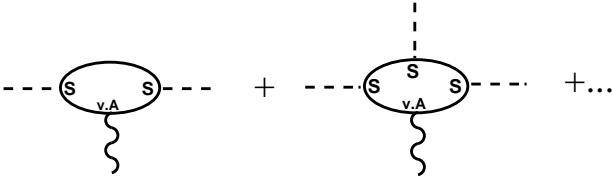}}
\caption{Spin response to a transverse magnetic field ${\bf B}_{\perp}$ in the presence of a current: perturbation theory to all orders in ${\bf B}_{\perp}$. 
The quasiparticles (quasiholes) in these diagrams diagonalize a Hamiltonian whose exchange field is pointing 
along an easy direction and ${\bf B}_{\perp}$ is by definition perpendicular to this easy direction. 
Provided that in Eq.~(\ref{eq:H_an_eq2}) we take the exact eigenstates of the mean field Hamiltonian (within which the exchange field need not be pointing along an easy direction), all the diagrams of this figure are implicit in the diagram of Fig.~(\ref{fig:bubbles}). In particular, the ferromagnetic ISGE captures the influence of currents on ferromagnetic resonance.\label{fig:three}} 
\end{center}
\end{figure}

Eqs.~(\ref{eq:occupation}) and~(\ref{eq:state2}) highlight the two ways in which a current alters the magnetic anisotropy field. Eq.~(\ref{eq:occupation}) captures the shift in the 
effective quasiparticle energies due to acceleration by an electric field, while Eq.~(\ref{eq:state}) describes the modification of the quasiparticle wavefunctions. 
As will become clear below the former is associated with intraband contributions to the anisotropy field
 whereas the latter may be traced to the interband contributions. Interband contributions are often neglected \cite{manchon,liu} because they are parametrically smaller by a factor of scattering rate $\tau^{-1}$ in good conductors.
However, as we show in the next section they may become quantitatively significant in disordered magnets like the (III,Mn)V materials.\cite{jungwirth}
Admittedly, other corrections with the same parametric dependence on disorder strength could also be present - but the 
description of these would require 
a detailed characterization of the disorder potential and a more
sophisticated transport theory.  The effect we retain is analogous to the intrinsic contribution to the anomalous Hall effect.\cite{nagaosa}  
Substituting Eqs.~(\ref{eq:occupation}), ~(\ref{eq:state}) and ~(\ref{eq:state2}) in Eq.~(\ref{eq:d_H_an}) we obtain 
\begin{equation}
{\bf \delta H}_{\rm an}={\bf \delta H}_{\rm an}^{\rm intra}+{\bf \delta H}_{\rm an}^{\rm inter}\nonumber\\
\end{equation}
where
\begin{eqnarray}
{\bf \delta H}_{\rm an}^{\rm intra}&=&\frac{1}{S_0}\sum_{{\bf k},a} [\Delta_0({\bf r}){\bf s}]_{a,a}{\bf v}_{a,a}\cdot{\bf E} \, \frac{\partial f_{{\bf k},a}}{\partial\epsilon_{{\bf k},a}} \, \tau_{{\bf k},a}\nonumber\\
{\bf \delta H}_{\rm an}^{\rm inter}&=&\frac{1}{i\omega}\frac{1}{S_0}\sum_{{\bf k},a\neq b} [\Delta_0({\bf r}){\bf s}]_{a,b}{\bf v}_{b,a}\cdot{\bf E} \;\nonumber\\
&&~~~~~~~~~~~~~~\times\frac{f_{{\bf k},a}-f_{{\bf k},b}}{\epsilon_{{\bf k},b}-\epsilon_{{\bf k},a}+\omega+i\eta}
\end{eqnarray}
In the expression for ${\bf \delta H}_{\rm an}^{\rm inter}$ we have selected the coefficient of $\exp(i\omega t)$ in the perturbation expansion, have neglected disorder scattering and have allowed for a small positive imaginary part in the frequency.   

Several remarks are pertinent in regards to our derivation of the interband component. First, it should be noted that in the zero frequency limit the imaginary part of ${\bf\delta H}_{\rm an}^{\rm inter}$ 
gets cancelled by the diamagnetic contribution, in such a way that the anisotropy field induced by a DC current is finite and real. 
Second, it is instructive to elaborate on the real part of ${\bf\delta H}_{\rm an}^{\rm inter}$:
\begin{eqnarray}
\label{eq:H_inter}
&&\delta {\bf H}_{\rm an}^{\rm inter}=\nonumber\\ 
&&=\frac{-\pi}{S_0\omega}\sum_{{\bf k},a\neq b} {\rm Re}\left[(\Delta_0{\bf s})_{a,b}{\bf v}_{b,a}\right]\cdot{\bf E}(f_{{\bf k},a}-f_{{\bf k},b})\delta(\omega_{b,a}+\omega)\nonumber\\
&&+\frac{1}{S_0\omega}\sum_{{\bf k},a\neq b} {\rm Im}\left[(\Delta_0{\bf s})_{a,b} {\bf v}_{b,a}\right]\cdot{\bf E}f_{{\bf k},a} \frac{2\omega}{\omega^2-\omega_{b,a}^2},
\end{eqnarray}
where $\omega_{b,a}\equiv\epsilon_{{\bf k},b}-\epsilon_{{\bf k},a}$. >From Eq.~(\ref{eq:H_inter}) it is clear that $\delta{\bf H}_{\rm an}^{\rm inter}$ remains finite as $\omega\to 0$. 
When disorder is included in the above expressions, the contribution from the second line in Eq.~ (\ref{eq:H_inter}) scales as $\tau^{-1}$ and thus is unimportant when the broadening of the energy bands due to impurity scattering is small compared to the energy difference between states connected by interband transitions. In contrast, the third line scales as $\tau^0$, and therefore it supplies the bulk of the interband contribution in weakly disordered ferromagnets. 

Recognizing the fact that the integration of equal-band Green's functions gives rise to a factor of $\tau$, ${\bf\delta H}_{\rm an}^{\rm intra}$ yields the intraband piece of Eq.~(\ref{eq:chi}) modulo a factor of $\Delta_0/S_0$. Similarly, ${\bf\delta H}_{\rm an}^{\rm inter}$ brings in the interband part of Eq.~(\ref{eq:chi}) modulo a factor of $\Delta_0/S_0$; in order to verify this we recall\cite{disorder} that
\begin{equation}
\label{eq:trick}
\sum_{{\bf k}}\frac{f_{{\bf k},a}-f_{{\bf k},b}}{\epsilon_{{\bf k},b}-\epsilon_{{\bf k},a}+i\omega}=-T \sum_{\omega_n,{\bf k}} G_a(i\omega_n,{\bf k})G_b(i\omega_n+i\omega,{\bf k}).
\end{equation}
In sum, we find
\begin{widetext}
\begin{equation}
\label{eq:complete}
\frac{\partial \delta H^i_{\rm an}}{\partial E^j}=\frac{1}{2\pi S_0}{\rm Re}\sum_{{\bf k},a,b} \langle a,{\bf k}|\Delta_0 ({\bf r}) s^i|b, {\bf k}\rangle\langle b,{\bf k}|v^j|a,{\bf k}\rangle\left(G_{{\bf k},a}^R G_{{\bf k},b}^A-G_{{\bf k},a}^R G_{{\bf k},b}^R\right),
\end{equation}
which agrees with the ISGE expression for the current-induced spin density (Eq.~(\ref{eq:chi})) except for an overall normalization factor ($1/S_0$) and the fact that the spin-operator is weighted by an spatially inhomogeneous magnitude of the exchange field. With the aim of making Eq.~(\ref{eq:complete}) more manageable for first principles calculations, we will ignore the interband contribution as well as the $G^R G^R$ term; both omissions are justified in most metallic ferromagnets.\cite{ahe term} In this case Eq.~(\ref{eq:complete}) simplifies into
\begin{equation}
\label{eq:simplified}
\frac{\partial \delta H^i_{\rm an}}{\partial E^j}\simeq\frac{1}{S_0}\sum_{{\bf k},a} \langle a,{\bf k}|\frac{{\partial\cal H}_{\rm so}}{\partial\Omega_i}|a, {\bf k}\rangle\langle a,{\bf k}|v^j|a,{\bf k}\rangle\frac{\partial f_{{\bf k},a}(\hat\Omega)}{\partial\epsilon_{{\bf k},a}}\tau_{{\bf k},a},
\end{equation}
\end{widetext}
where we have re-inserted $\langle a|\Delta_0({\bf r}) {\bf s}|a\rangle =\langle a|\partial{\cal H}_{\rm so}/\partial\hat\Omega| a\rangle$. 
 While approximate, Eq.~(\ref{eq:simplified}) may provide a valid platform to explore current induced magnetization reversal in real gyrotropic ferromagnets with a single magnetic domain. In the next section we will describe in detail how a large  $\delta H_{\rm an}$ can produce a large reorientation of the magnetization. 

If the spatial dependence of $\Delta_0({\bf r})$ is negligible (as it will be in the model studied in the next section), Eq.~(\ref{eq:complete}) may be rewritten in a more compact way:
\begin{equation}
\label{eq:link}
\chi^{i,j}_{S,E}=\frac{S_0}{\Delta_0}\frac{\partial \delta H^i_{\rm an}}{\partial E^j}.
\end{equation}
where $\chi_{S,E}$ is the spin-current susceptibility introduced in Eq.~(\ref{eq:chi}). Eq.~(\ref{eq:link}) proves that the ferromagnetic ISGE describes the change in the magnetic anisotropy field due to a current. 
In other words, ferromagnetic ISGE determines how an electric current changes the location of the extrema in the micromagnetic energy functional. This is the central idea of this section.



As a final sidenote, we point out that this section has concentrated on evaluating the change in magnetic anisotropy under a perturbation represented by ${\bf v}\cdot{\bf A}$, where ${\bf A}$ is the electromagnetic vector potential.  The anisotropy is evaluated by calculating the change in the expectation value of $\Delta_0 s$, thus leading to a 
rather standard linear response function calculation.  We could in the same way calculate the change in the transverse spin-spin response function due to 
an electric field as indicated in Fig.(~\ref{fig:three}), in order to determine how small amplitude magnetic fluctuations are altered.
If, however, we are interested only in uniform magnetization dynamics no additional 
information is obtained by doing this calculation.  The key point
is that the response to a transverse field ${\bf B}_{\perp}$  is already built in our expression for the equilibrium anisotropy field (Eq.~(\ref{eq:H_an_eq2})), to {\em all} orders in ${\bf B}_{\perp}$. 
In other words, the reference (unperturbed) macrostate to which we apply a current contains a magnetization that is ``arbitrarily'' misaligned with the exchange field. Hence, Eq.~(\ref{eq:link}) along with Eq.~(\ref{eq:H_an_eq2}) offers a complete account of the nonequilibrium magnetic anisotropy of uniform magnetic states in the presence of a transport current. 

\section{Current-Driven Magnetization Reversal in Monodomain (Ga,Mn)As}

Magnetoelectric phenomena in dilute magnetic semiconductors\cite{jungwirth} such as (Ga,Mn)As
have attracted special attention because these materials are 
more compatible with current microelectronics technology than metals.  
In addition, electric field control of magnetism has turned out to be more feasible in (Ga,Mn)As than in conventional dense-moment metallic ferromagnets because of their 
small magnetization, 
high carrier spin polarization, strong spin-orbit interactions, and carrier-mediated ferromagnetism.\cite{chiba,stolichnov, ohno_dietl} 
In particular, the recent experiment\cite{chernyshov} by Chernyshov {\em et al.} on (Ga,Mn)As wafers under compressive strain has
demonstrated the ability of transport currents to reversibly {\em assist} the reorientation of magnetization in {\em single-domain} ferromagnets.
As we demonstrate here this effect is dependent on having both spin-orbit interactions and broken inversion symmetry. 
In this section we compute the change in the magnetic anisotropy due to an electric current for a realistic model of (Ga,Mn)As.
Our calculation is directly relevant to the experiment of Chernyshov {\em et al.}. Our results corroborate their interpretation of the data and predict the possibility of all-electric magnetization switching in (Ga,Mn)As. Our analysis is limited to zero temperature and neglects the shape anisotropy, which for typical Mn doping concentrations is 10-100 times weaker than in conventional ferromagnets.

\begin{figure}
\begin{center}
\scalebox{0.4}{\includegraphics{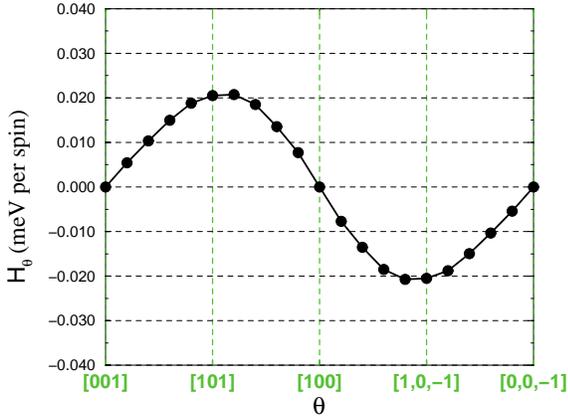}}
\caption{Equilibrium anisotropy field (meV per spin) in (Ga,Mn)As for $\phi=0$, and $\theta\in(0,\pi)$. The parameters 
used for this calculation were: Mn fraction $x=0.08$, hole concentration $p\simeq 0.15 {\rm nm}^{-3}$, $\epsilon_F\tau=3$, and 
axial strain $\epsilon_{\rm ax}=-0.5\%$.
These anisotropy field results were evaluated using the model explained in the text.}
\label{fig:H_an_eq1}
\end{center}
\end{figure}

\begin{figure}
\begin{center}
\scalebox{0.4}{\includegraphics{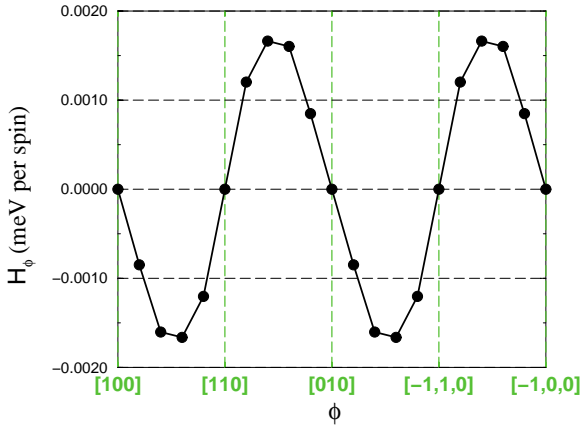}}
\caption{Equilibrium anisotropy field (meV per spin) in (Ga,Mn)As for $\theta=\pi/2$ and $\phi\in(0,\pi)$.
The parameters are: Mn fraction $x=0.08$, hole concentration $p\simeq 0.15 {\rm nm}^{-3}$, $\epsilon_F\tau=3$, and axial strain $\epsilon_{\rm ax}=-0.5\%$. 
These results were evaluated using the model explained in the text. Due to strain, the in-plane anisotropy is notably weaker than the out-of-plane anisotropy represented in the previous figure.}
\label{fig:H_an_eq2}
\end{center}
\end{figure}

The dependence of the magnetic anisotropy of (Ga,Mn)As on doping, external electric fields, temperature and strain has been successfully 
explained\cite{abolfath,dietl,jan} by combining (i) a mean-field theory of the exchange coupling between localized Mn moments and  valence band carriers with (ii) a phenomenological four or six band envelope function model in which the valence band holes are characterized by Luttinger, spin-orbit splitting and strain-energy parameters.
The results presented below predict the rate at which these fields change with external electric field.

\begin{figure}
\begin{center}
\scalebox{0.4}{\includegraphics{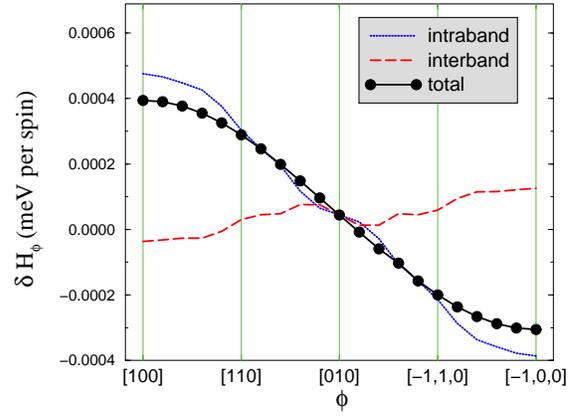}}
\caption{Change in the magnetic anisotropy field of (Ga,Mn)As (in meV per spin) due to the inverse spin-galvanic effect, for an electric field of $1 {\rm mV/nm}$ along [010]. The parameters are: Mn fraction $x=0.08$, hole concentration $\simeq 0.25 {\rm nm}^{-3}$, $\epsilon_F\tau=2$, and axial strain $\epsilon_{\rm ax}=-1\%$ 
We compare between interband and intraband contributions: in contrast to the case of good metals, the 
interband contributions are not negligible in (Ga,Mn)As.  
For the present case, had we neglected the interband contribution the minimum electric field needed to reorient the magnetization by 90$^\circ$ would be off by approximately 20 \%. 
The sum of interband and intraband pieces gives rise to a smooth curve that reflects the Dresselhaus
symmetry of the axial strain. Reversing the sign of the axial strain (i.e. making it tensile) leads to a sign reversal of $\delta H_{\phi}$.}
\label{fig:intra vs inter}
\end{center}
\end{figure}

\begin{figure}
\begin{center}
\scalebox{0.4}{\includegraphics{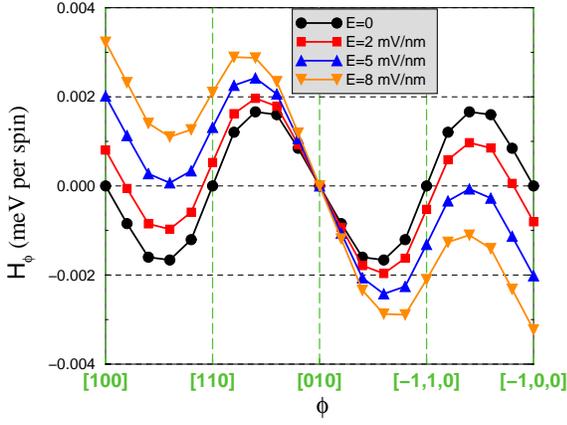}}
\caption{Reorientation of the magnetization due to an electric current. An initial magnetization along [100] can be rotated (assisted by damping) into [010] by applying a sufficiently strong electric field with a nonzero projection along the [010] direction (a current along [100] would not destabilize the [100] easy axis). For the parameters of this figure ($x=0.08$, $p\simeq 0.15 nm^{-3}$, $\epsilon_F\tau=3$, $\epsilon_{\rm ax}=-0.5\%$) the critical electric field is $\simeq 5 {\rm mV/nm}$, which corresponds roughly to a critical current density of $5 \times 10^7 {\rm A/cm}^2 $.}
\label{fig:H_an_neq1}
\end{center}
\end{figure}

In line with this  we adopt the following Hamiltonian for Ga$_{1-x}$Mn$_x$As:
\begin{equation}
\label{eq:model}
{\cal H}=\cal{H}_{\rm KL}+\cal{H}_{\rm strain}+{\bf S}\cdot{\bf \Delta}.
\end{equation}
$\cal{H}_{\rm KL}$ is the 4-band Kohn-Luttinger Hamiltonian\cite{cardona} with Luttinger parameters $\gamma_1=6.98$, $\gamma_2=2.1$ and $\gamma_3=2.9$. ${\bf S}$ is the spin operator projected onto the J=3/2 total angular momentum subspace at the top of the valence band. ${\bf \Delta}=\Delta_0 \hat\Omega = J_{\rm pd} S N_{\rm Mn}\hat\Omega$ is the exchange field, $\hat\Omega$ denotes the orientation of the local moments, $J_{\rm pd}=55\mbox{ meV nm}$ is the p-d exchange coupling parameter, $S=5/2$ is the spin of the Mn ions, and $N_{\rm Mn}=4 x/a^3$ is the Mn concentration ($a=0.565\mbox{nm}$ is the lattice constant of GaAs).   
This four-band model is expected to be adequate for small and intermediate Mn doping strengths.  ${\cal H}_{\rm strain}$ is the strain Hamiltonian\cite{chernyshov, silver, winkler} given by
\begin{eqnarray}
\label{eq:strain}
{\cal H}_{\rm strain}&=& -b\left[ \left(J_x^2-\frac{{\bf J}^2}{3}\right)\epsilon_{xx}+ c.p.\right]\nonumber\\ 
&+& C_4 \left[ J_x \left(\epsilon_{yy}-\epsilon_{zz}\right) k_x + c.p.\right],
\end{eqnarray}
where ${\bf J}$ is the total angular momentum (${\bf J}=3{\bf S}$ by the Wigner-Eckart theorem), $\epsilon_{i,i}$ are diagonal elements of the
stress tensor, 
$b=-1.7\mbox{ eV}$ is the axial deformation potential and
the parameter $C_4=5\mbox{ eV \AA}$ captures the strain-induced linear in $k$ spin-splitting of the valence bands in paramagnetic GaAs.
In Eq.~(\ref{eq:strain}) the notation {\em c.p.} stands for cyclic permutations and $\epsilon_{x,x}=\epsilon_{y,y}\neq\epsilon_{z,z}$ 
for $[001]$ 
growth lattice-matching strains. The term proportional to $C_4$ is crucial for the occurrence of the 
ferromagnetic ISGE because it breaks inversion symmetry (we are neglecting the intrinsic lack of inversion symmetry of the zinc-blende structure, which is relatively inconsequential), 
and it introduces chirality.  (A bulk, unstrained zinc-blende crystal is not gyrotropic because it corresponds to the $T_d$ symmetry point group.)
Eq.~(\ref{eq:strain}) may be simplified to
\begin{equation}
{\cal H}_{\rm strain}=-b \epsilon_{\rm ax}\left(J_z^2-\frac{{\bf J}^2}{3}\right)+C_4\epsilon_{\rm ax}\left(J_y k_y-J_x k_x\right),
\end{equation}
where $\epsilon_{\rm ax}=\epsilon_{zz}-\epsilon_{xx}$ is the purely axial strain component. In this paper we take $\epsilon_{\rm ax}<0$ (compressive strain), which applies when (Ga,Mn)As is grown on top of a GaAs substrate.

\begin{figure}
\begin{center}
\scalebox{0.4}{\includegraphics{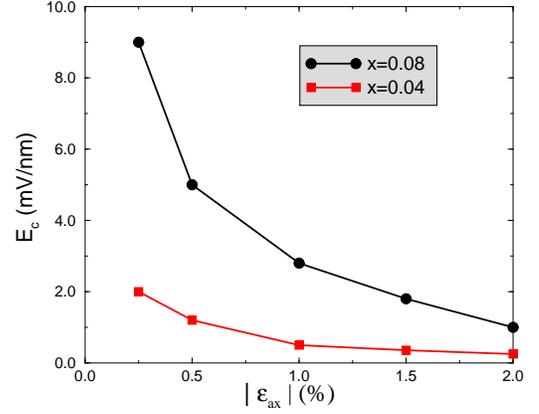}}
\caption{Dependence of the critical electric field (at which the magnetization gets reoriented by 90$^\circ$) on (compressive) axial strain. 
The critical current is (roughly) inversely proportional to $\epsilon_{\rm ax}$.  The reason behind this relationship is that the equilibrium, {\em azimuthal} anisotropy is largely indiferent to $\epsilon_{\rm ax}$. For $x=0.04$ and $\epsilon_{\rm ax}=-2\%$ we find $E_c\simeq 0.25 {\rm mV/nm}$, which corresponds to a critical current on the order of $10^6 {\rm A/cm^2}$. These results are for a (Ga,Mn)As model
with carrier density $p\simeq 0.15 nm^{-3}$ and $\epsilon_F\tau=3$.} 
\label{fig:critical}
\end{center}
\end{figure}

Using Eqs.~(\ref{eq:H_an_eq2}) , ~(\ref{eq:link}) and ~(\ref{eq:model}) we evaluate the magnetic anisotropy field both with and without electric current; the results are highlighted in Figs. ~(\ref{fig:H_an_eq1})-~(\ref{fig:critical}). 
Figs.~(\ref{fig:H_an_eq1}) and ~(\ref{fig:H_an_eq2}) correspond to electrical equilibrium and illustrate $H_{\theta}=-1/S_0\sum_{{\bf k},a} (\partial\epsilon_{{\bf k},a}/\partial\theta) f_{{\bf k},a}$ for $\phi=0$ and $H_{\phi}=-1/S_0\sum_{{\bf k},a} (\partial\epsilon_{{\bf k},a}/\partial\phi) f_{{\bf k},a}$ for $\theta=\pi/2$, respectively. 
The extrema of the micromagnetic energy functional are characterized by $H_{\phi}=H_{\theta}=0$ and by inspection we locate them at $\theta=0$ and  $(\theta,\phi)=(\pi/2,n \pi/4)$ where $n=0,1,2...$. 
For our parameters (see figure captions) the energy minima that define metastable magnetic configurations are found at $(\theta,\phi)=(\pi/2,n \pi/2)$.
That is to say, the easy directions correspond to [100], [010],[$\bar{1}$00] and [0$\bar{1}$0], which are contained in the plane of the (Ga,Mn)As wafer. 
For later reference, we consider an initial condition in which the magnetization is pointing along [100]. If a small static perturbation tilts it towards [110], the negative anisotropy field ($H_{\phi}<0$ for $\phi\gtrsim 0$) creates a torque that will, in conjunction with damping,\cite{damping} turn the magnetization back to [100]. 

Fig. ~(\ref{fig:intra vs inter}) illustrates how an electric current along [010] alters the azimuthal anisotropy field\cite{caveat} for fixed $\theta=\pi/2$. The cosine-like shape is consistent with the Dresselhaus symmetry of the $C_4$ term in the strain Hamiltonian. If the system had a perfect Dresselhaus symmetry the change in the micromagnetic energy functional under an electric current ${\bf j}$ would read 
\begin{equation}
\label{eq:dress}
\delta E_{GS}\propto C_4 \epsilon_{\rm ax}(\Omega_y j_y - \Omega_x j_x), 
\end{equation}
which means that a current along [010] ([100]) would tilt the steady-state magnetization
direction along [010] ([${\bar 1}$00]).
Using $\Omega_x=\sin\theta\cos\phi$ and $\Omega_y=\sin\theta\sin\phi$ it follows that $\delta H_{\phi}\propto j_y \cos\phi + j_x \sin\phi$, and hence a cosine-like dependence in $\phi$ is indeed expected for a current along [010].  We have verified that a current along ${\bf x}$ gives rise to a sine-like dependence with the appropriate sign. 
Nevertheless, Eq.~ (\ref{eq:dress}) is not exact because the magnetization vector introduces another preferred direction; for instance, we find that an electric field pointing along ${\hat z}$ (i.e. [001]) can also alter the steady-state spin orientation. This effect, which vanishes in the paramagnetic limit, highlights one instance in which the 
ferromagnetic and paramagnetic ISGEs differ.
Another attribute of Fig. ~(\ref{fig:intra vs inter}) is that it determines the quantitative importance of interband contributions to the current-induced spin density in (Ga,Mn)As.  Although normally neglected, interband transitions become quantitatively significant in strongly disordered ferromagnets. In particular, interband and intraband contributions are largely indistinguishable in ferromagnets with $\Delta_0\tau< 1$.  We note parenthetically that neither intraband nor interband contributions display the smooth sinusoidal shape portrayed by their sum. In addition, we remark that reversing the sign of the axial strain (i.e. making it tensile) leads to a sign reversal of $\delta H_{\phi}$ without substantial changes in its magnitude.\cite{quantum well}

Fig.~(\ref{fig:H_an_neq1}) demonstrates that a sufficiently
strong current is able to rotate the magnetization by $90^\circ$ or $180^\circ$.
We explain this property by considering the case in which the equilibrium magnetization is pointing along [100]. 
If a small current is applied along [010], then [100] is no longer an extremum of the micromagnetic energy functional (because $H_{\phi}(\phi=0)\propto E_y\neq 0$).
The modified {\em easy} direction remains in the neighborhood of [100] since the restoring torque ($H_{\phi}<0$) again crosses zero at $\phi\gtrsim 0$. 
Once the applied electric field exceeds a critical value ($E_c\simeq {\rm 5.5 mV/nm}$ in the present figure) the $H_{\phi}<0$ region near [100] disappears completely and hence assisted by damping the magnetization eventually 
points along [010]. In other words, at (and above) the critical switching field the energy minimum that is nearest to [100] is located at [010] (note that this direction remains stable when the current flows along [010]). 
Once the magnetization is aligned with [010], an equally strong electric current in the [100] direction will rotate it towards [$\bar{1}$00]. 
In this fashion it is possible to switch the direction of magnetization by $180^\circ$ solely by application of transport currents. 

The procedure sketched above accomplishes magnetization switching by application of two perpendicular current pulses, each of which forces a $90^\circ$ rotation. Yet, it is also possible to achieve the $[100]\to[{\bar 1}00]$ switching with a single {\em unidirectional} pulse, provided the electric field along [100] is ramped up sufficiently ($E_{c,2}\simeq 20 {\rm mV/nm}$ for the parameters of the present figure). 
In order to understand this, recall that ${\bf j}||{\bf\hat x}\rightarrow \delta {\bf H}_{\rm an}||-{\bf\hat x}$. Consequently, for a strong electric current [${\bar 1}$00] is the only easy direction ([100] becomes a hard direction). The inequivalence between [100] and [$\bar{1}$00] does not violate any symmetry principles;\cite{stohr latest} in effect, an electric current breaks time reversal symmetry and can thus connect time-reversed magnetic states.

Using $\rho=10^{-3} \Omega {\rm cm}$ as the typical resistivity for (Ga,Mn)As samples we deduce that $E=1 {\rm mV/nm}$ 
corresponds approximately to a current density of $10^7 {\rm A/cm^2}$, hence the critical switching current is on the order of $10^6-10^7 {\rm A/cm^2}$. 
It is plausible that a detailed exploration of the parameter space comprised by the Mn concentration $x$, the hole density $p$ and the axial strain $\epsilon_{\rm ax}$ will enable lower critical currents, thereby diminishing the importance of the Joule heating.  
As a word of caution, we note that the 4-band model employed here typically overestimates the effect of spin-orbit interactions, 
thus potentially leading to an underestimate of these critical currents.  There is in addition some uncertainty associated with 
the use of a life-time approximation for Bloch state quasiparticles in these strongly disordered metallic conducting ferromagnets.

Overall, the magnitude of the critical switching current depends on (a) the size of the equilibrium anisotropy barrier, (b) the extent to which inversion symmetry is broken and (c) the strength of spin-orbit interaction. 
In (Ga,Mn)As the first two factors are tunable. On one hand, (a) may be optimized by choosing appropriate doping concentrations: in general lower Mn density is beneficial (Fig.~(\ref{fig:critical})), as it reduces the equilibrium anisotropy without significantly affecting the magnitude of ISGE. However, for very low Mn concentrations a metal-insulator transition is impending, which hampers ISGE.
On the other hand, (b) may be modified via strain engineering: as shown in Fig. ~(\ref{fig:critical}), the critical current is (roughly) inversely proportional to the strength of the uniaxial strain that breaks inversion symmetry. The inverse proportionality may be understood on the basis of Eq.~(\ref{eq:dress}) combined with the fact that the equilibrium anisotropy does not change to {\em first} order in $\epsilon_{\rm ax}$ (because ${\bf k}$-linear terms vanish after summing over all momenta).

\section{Summary and Conclusions}

In this work we have presented a theory of the current-induced spin torques in uniform ferromagnets.
The torques can be viewed as due to a difference between the magnetic anisotropy energy of 
a ferromagnet which carries no current and the magnetic anisotropy of a ferromagnet 
in the transport steady state, which give rise to a corresponding change in 
anisotropy effective magnetic fields.  When the transport steady state is described using a relaxation time 
approximation, the current-induced contribution to the anisotropy field of a strongly metallic ferromagnet is given 
in energy units by 
\begin{equation}
{\bf \delta H}_{\rm an}= \frac{1}{S_0}\sum_{{\bf k},a} [\Delta_0({\bf r}){\bf s}]_{a,a}{\bf v}_{a,a}\cdot{\bf E} \, \frac{\partial f_{{\bf k},a}}{\partial\epsilon_{{\bf k},a}} \, \tau_{{\bf k},a}.
\end{equation} 
where $[\Delta_0({\bf r}){\bf s}]_{a,a}$ is the spin-density weighted average of the exchange splitting of a particular state.
We refer to the existence of this current-induced anisotropy field as the ferromagnetic inverse spin-galvanic effect.

In bulk materials this current induced field is non-zero only in gyrotropic ferromagnets, {\em i.e.} only 
in ferromagnets that 
have broken inversion symmetry and are chiral.   
Although uniform ferromagnetism may appear to be incompatible with broken inversion symmetry because of the  
the Dzyaloshinskii-Moriya interaction,
the equilibrium magnetic anisotropy is often strong enough
(or at least can be engineered so that it is strong enough)
to prevent the formation of spiral magnetic states. 

As an illustration of our theory, we have estimated current induced
torques in uniform (Ga,Mn)As, which is not gyrotropic when it has pseudo-cubic
symmetry but becomes gyrotropic when strained.  Since substrate-dependent strains are
present in all (Ga,Mn)As thin films, the strength of the ferromagnetic 
ISGE is expected to be strongly sample-dependent. We have concluded that it should {\em a priori} 
be feasible to design (Ga,Mn)As samples in which it is possible to switch the magnetization purely by electrical means. 
For typical sample parameters the necessary switching currents are on the order of $10^6-10^{7} {\rm A/cm^2}$, 
but the value may be tuned by adjusting the doping concentration and the axial strain. 
At these critical currents the Joule heating is not negligible; however, it is possible that further studies exploring the entire parameter space 
of Mn concentration,  hole density, and the axial strain will identify circumstances under which the critical currents are smaller.

Another possible avenue for further research consists of evaluating the anisotropy fields which can be generated 
by electrical currents in strain engineered samples of appropriate technologically useful ferromagnets.
Since we are not aware of room-temperature transition metal ferromagnets that are gyrotropic,\cite{krupin} we propose arranging a room-temperature, non-gyrotropic ferromagnet (e.g. permalloy) in contact with a non-magnetic, gyrotropic material (e.g. strained GaAs). 
In these artificial heterostructures room-temperature magnetism and gyrotropic symmetry would coexist by virtue of the proximity effect. 

Finally, effects similar to those studied in this work would allow transport currents to change spiral 
states, and possibly to induce or remove them.

\acknowledgements
The authors thank S. Ganichev, T. Jungwirth and J. Stohr for helpful advice.
This work was supported by the Welch Foundation and by the National Science Foundation under grant 
DMR-0606489.


\begin{thebibliography}{30}
\bibitem{inversion} All gyrotropic crystals are non-centrosymmetric, but not all non-centrosymmetric crystals are gyrotropic.
\bibitem{symmetry arguments} For paramagnetic metals it is straightforward to deduce the relative orientation between the applied electric field and the induced spin polarization from symmetry arguments. 
For instance, the Hamiltonian for a 2DEG with Rashba spin-orbit interaction is invariant under a 90$^\circ$ rotation around the $\hat z$ as well as under a reflection with respect to the plane perpendicular to $\hat y$. 
According to Neumann's principle, the tensor relating the electric field and the spin polarization must be invariant under the above mentioned symmetry operations. 
From this requirement it follows that the spin polarization must be perpendicular to the current.
\bibitem{diamagnetic} In the zero frequency limit the real part of $\tilde\chi/\omega$ is cancelled out by the diamagnetic response. This reflects the fact that in non-superconducting metals the current induced spin density is dissipative.
\bibitem{disorder} Strictly speaking Eqs. ~(\ref{eq:state2}) and ~(\ref{eq:trick}) are accurate only in absence of disorder. Nevertheless, the connection between them remains intact in presence of impurities.
\bibitem{ahe term} The leading O($\tau^0$) correction due to interband transitions would be captured by the second line of Eq.~(\ref{eq:H_inter}).
\bibitem{damping} In absence of damping the magnetization would keep precessing indefinitely. The combined action of the anisotropy field and damping is what ultimately drives the system to the minimum energy state.
\bibitem{caveat} Our discussion concentrates on $H_{\phi}$. Although $H_{\theta}$ too generally changes in presence of a current, it is not pertinent to the [100]$\to$[010] or [100]$\to$[${\bar 1}$00] magnetization reorientations that we are interested in.
\end{thebibliography}
\end{document}